\documentclass[]{article}

\usepackage{graphicx}

\begin{document}

\begin{center}

{\bf Laboratory Limits on Temporal Variations of Fundamental Constants:\\ An Update}\\

\bigskip

{\sc E. Peik, B. Lipphardt, H. Schnatz, Chr. Tamm,\\ S. Weyers, R. Wynands}

\bigskip

{\it Physikalisch-Technische Bundesanstalt,\\
Bundesallee 100, 38116 Braunschweig, Germany}

\end{center}

\vspace{1cm}

\begin{abstract}
Precision comparisons of different atomic frequency standards over a period of a few years can be used for a sensitive search for temporal variations of fundamental constants. 
We present recent frequency measurements of the 688 THz transition in the $^{171}$Yb$^+$ ion. For this transition frequency a record over six years is now available, showing that a possible frequency drift relative to a cesium clock can be constrained to $(-0.54\pm0.97)$~Hz/yr, i.e. at the level of $2\cdot10^{-15}$ per year. Combined with precision frequency measurements of an optical frequency in $^{199}$Hg$^+$ and of the hyperfine ground state splitting in $^{87}$Rb a stringent limit on temporal variations of the fine structure constant $\alpha$: 
$d\ln\alpha /dt= (-0.26\pm0.39)\cdot 10^{-15}~{\rm yr}^{-1}$ 
and a model-dependent limit for variations of the proton-to-electron mass ratio $\mu$ in the present epoch can be derived: 
$d \ln \mu/dt = (-1.2 \pm 2.2)\cdot 10^{-15}~{\rm yr}^{-1}$.
We discuss these results in the context of astrophysical observations that apparently indicate changes in both of these constants over the last 5--10 billion years.  
\end{abstract}

\vspace{1cm}

\noindent
{\bf 1. Search for variations of constants in astrophysics and with atomic clocks}
\medskip

Over the past few years there has been great interest in the possibility that the fundamental constants of nature might show temporal variations over cosmological time scales \cite{uzan,acfc,barrow,karsh,bronn}. Such an effect -- as incompatible as it seems with the present foundations of physics -- appears quite naturally in the attempt to find a unified theory of the fundamental interactions. An active search for an indication of variable constants is pursued mainly in two areas: observational astrophysics and laboratory experiments with atomic frequency standards.   
The two most important quantities under test are Sommerfeld's fine structure constant $\alpha=e^2/(4\pi\epsilon_0\hbar c)$ and the proton-to-electron mass ratio $\mu=m_p/m_e$. While $\alpha$ is the coupling constant of the electromagnetic interaction, $\mu$ also depends on the strength of the strong force and on the quark masses via the proton mass. 
It is important that these quantities are dimensionless numbers so that results can be interpreted independently from the conventions of a specific system of units. Quite conveniently, both constants appear prominently in atomic and molecular transition energies: $\alpha$ in atomic fine structure splittings and other relativistic contributions, and $\mu$ in molecular vibration and rotation frequencies as well as in hyperfine structure.

A multitude of data on variations of $\alpha$ and $\mu$ has been obtained from astrophysical observations but the present picture that is obtained is not completely consistent: 
Evidence for a variation of $\alpha$ has been derived from a shift of wavelengths of metal ion absorption lines produced by interstellar clouds in the light from distant quasars \cite{webb1}. These observations
suggest that about 10 billion years ago (redshift range $0.5<z<3.5$),
the value of $\alpha$ was smaller than today by $\Delta \alpha / \alpha= (-0.543\pm 0.116)\cdot 10^{-5}$, 
representing  $4.7\sigma$ evidence for a varying $\alpha$ \cite{webb2}. (Here the definition is used that a negative sign of $\Delta\alpha$ corresponds to a temporal increase of the constant.) 
Assuming a linear increase of $\alpha$ with time, this would correspond to a drift rate
$d \ln \alpha / d t=(6.40\pm 1.35)\cdot 10^{-16}~{\rm yr}^{-1}$ \cite{webb2}.
Other evaluations of spectra obtained with a different telescope and using other selections of quasar absorption systems reach similar or even higher sensitivity for $\Delta \alpha / \alpha$ but are consistent with $\Delta \alpha=0$  for all look-back times \cite{quast,sria}. 

In the case of the proton-to-electron mass ratio a relative change
$\Delta \mu/\mu = (2.4\pm 0.6)\cdot 10^{-5}$ has been inferred from the analysis of absorption systems of  H$_2$ clouds  at redshifts
$z = 2.6$ and $3.0$ (look-back time $\approx12$~Gyr), observed in the spectra of two quasars \cite{reinhold}. Assuming a linear time evolution, this would correspond to $d \ln \mu / d t\approx -2\cdot 10^{-15}~{\rm yr}^{-1}$. Looking only at the positive reports about variations, one would conclude $|\Delta \mu/\mu |> |\Delta \alpha/\alpha|$. Indeed, in the framework of theories of unification it is plausible that the scale of the strong interaction shows a bigger rate of change than the electromagnetic force \cite{marciano,fritzsch,langacker}. 

Other recent reports derive stringent limits on variations: an analysis of 18~cm OH lines at red shift $z=0.765$ (look-back time 6.7 billion years) constrains the variation of the quantity $F \equiv g_p(\alpha^2\mu)^{1.57}$ ($g_p$: $g$-factor of the proton) to 
$\Delta F/F= (0.44\pm 1.07)\cdot 10^{-5}$ \cite{kanekar}. A combination of the H ground state hyperfine splitting interval (21~cm line) and metal ion absorption lines in quasar spectra can be used to test the constancy of the parameter $X \equiv \alpha^2 g_p/\mu$ and a result $\Delta X/X=(1.17\pm1.01)\cdot 10^{-5}$ has been obtained for the redshift range $0.2<z<2.0$.\cite{tzanavaris} Combined with either the non-zero \cite{webb2} or the zero \cite{sria} value for $\Delta \alpha/\alpha$ one obtains  $\Delta \mu/\mu = (2.31\pm 1.03)\cdot 10^{-5}$ or $\Delta \mu/\mu = (1.29\pm 1.01)\cdot 10^{-5}$, respectively.
It should be kept in mind that the uncertainties cited here are usually the result of averaging over a large ensemble of data points with much higher individual statistical uncertainties. The significance of the results that detect non-zero variations therefore hinges on a complete understanding of all systematic effects that may introduce bias or correlations in the data and that may not be easy to detect.
     
The approach of using precision laboratory experiments with frequency standards to search for variations of the fundamental constants \cite{cjp} has the advantage that the relevant parameters are under the control of the experimenter, permitting a detailed investigation of possible systematic effects.
The obvious disadvantage is that the limited duration of the experiments of typically a few years only is not well adapted to search for evolution on a cosmological time scale. Application of the methods of laser cooling and trapping has led to significant improvements in the precision of atomic clocks and  frequency standards over the last years \cite{bauch,diddams}: Primary cesium clocks based on atomic fountains \cite{ww,bize} realize the SI second with a relative uncertainty below $1\cdot 10^{-15}$ and a similar level of accuracy has been achieved for the ground state hyperfine frequency of $^{87}$Rb \cite{bize}. Research towards {\em optical} clocks with trapped ions  and atoms \cite{gill,diddams} has resulted in several systems that approach the accuracy of primary cesium clocks or show an even higher reproducibility. 
In typical experiments performed so far to search for temporal changes of fundamental constants, an atomic transition frequency $f_{\rm at}$ is measured with respect to a cesium clock (i.e. the frequency ratio $f_{\rm at}/f_{\rm Cs}$ is determined, where $f_{\rm Cs}\equiv 9\,192\,631\,770$~Hz, the SI value of the ground state hyperfine splitting frequency of $^{133}$Cs). One measurement usually takes a few days of averaging time in order to reach a low statistical uncertainty. After a few years the measurement is repeated and changes in the frequency ratio can be looked for. 

This procedure has recently been undertaken for: the ground state hyperfine frequency of $^{87}$Rb \cite{marion,bize}, the $1S\rightarrow2S$ two-photon transition in atomic hydrogen \cite{fischer}, the transition $^2S_{1/2}\rightarrow {^2D_{5/2}}$ at 1065 THz in  Hg$^+$  \cite{udem2,oskay}, and the   $^2S_{1/2}\rightarrow {^2D}_{3/2}$ transition at 688~THz in $^{171}$Yb$^+$ (this work).
Significant changes of $\alpha$ or $\mu$ would produce a clear signature in these experiments, because they would influence the transition frequencies differently \cite{karsh,pk}. Atomic gross structure scales with the Rydberg constant $R_{\infty}$, hyperfine structure with the product of $\alpha^2 R_{\infty}$ and the nuclear magnetic moment. In addition,
the fine structure constant $\alpha$ appears in relativistic contributions to the level energies that increase with the square of the nuclear charge and are consequently much more important for heavy atoms \cite{prestage}. The sensitivity of a specific transition frequency to changes of $\alpha$ can be derived from {\em ab initio} relativistic atomic structure calculations \cite{dzu1,dzu2}.
While measurements on a single atomic gross structure transition relative to a cesium clock can test variations of a combination of $\alpha$ and the cesium magnetic moment only, a combination of data from different optical transition frequencies can be used to obtain a model-independent result about variations of $\alpha$ alone \cite{karsh,fischer,peik}. The published most stringent laboratory limit on variations of the fine structure constant in the present epoch had been obtained by this method in 2004:     
${d \ln \alpha}/{d t} = (- 0.3 \pm 2.0)\cdot 10^{-15}~{\rm yr}^{-1}$ \cite{peik}, using the three optical transitions mentioned above.
 
Contrary to the astrophysical observations, all present laboratory experiments are compatible with constancy of constants to within $1\sigma$.
The question arises to which degree the results of the quasar absorption spectra and the laboratory experiments can be interrelated, because they investigate very different time scales and different regions of the universe.
It has been pointed out recently that cosmological variations of physical constants linked to scalar fields should also be detectable on the surface of local gravitationally bound systems \cite{shaw}. This means that terrestrial laboratory experiments should in principle be able to reflect variations observed in interstellar gas clouds. Concerning the possible time dependence, one may expect the most dramatic variations during the very early phases of the evolution of the universe, e.g. during the inflationary period. 
Because the time scale on which variations seem to appear in the astrophysical observations is logarithmically much closer to our epoch than to the Big Bang, a simple linear time evolution of the constants may be used as a first approximation to model the behavior over the past few billion years.

In the following, we present recent precision measurements with the Yb$^+$ optical frequency standard at PTB that augment the existing record of data from this system. In combination with data from NIST, Boulder, on the Hg$^+$ standard a new model-independent limit on the variation of $\alpha$ is obtained. Combining this with data from LNE-SYRTE, Paris, on the constancy of the Rb/Cs hyperfine frequency ratio and using the Schmidt model for the nuclear $g$-factors of Rb and Cs, a limit on the variation of $\mu$ is obtained.   

\bigskip
\noindent
{\bf 2. Measurements of the Yb$^+$ optical frequency standard at 688~THz}
\medskip

The $^{171}$Yb$^+$ ion is attractive as an optical frequency standard because it offers narrow reference transitions with small systematic  frequency shift. At PTB, the electric quadrupole transition $(^2S_{1/2},F=0) \rightarrow (^2D_{3/2},F=2)$
at 436 nm wavelength (688 THz frequency) with a natural linewidth of 3.1 Hz is investigated \cite{tamm1,schneider,peik2,tamm2}.
A single ytterbium ion is trapped in a miniature Paul trap and is laser-cooled to a sub-millikelvin temperature
by exciting the low-frequency wing of the quasi-cyclic $(F=1) \rightarrow (F=0)$ component of the
$^2S_{1/2} \rightarrow {^2P_{1/2}}$ resonance transition at 370~nm (cf. Fig. 1). 
The $S\rightarrow D$ electric quadrupole reference transition is probed by the frequency doubled radiation from a diode laser
emitting at 871 nm. The short term frequency stability of this laser is derived from a temperature-stabilized and seismically isolated high-finesse reference cavity.

\begin{figure}[tbh]
\begin{center}
\includegraphics[width=9cm]{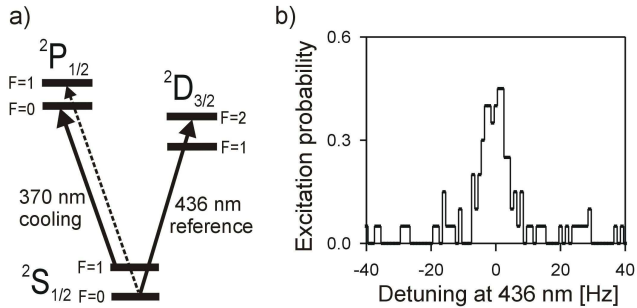}
\end{center}
\caption{a) Simplified level scheme of the $^{171}$Yb$^+$ ion. 
b) Excitation spectrum of the $^2S_{1/2}(F=0,m_F=0)\rightarrow {^2D}_{3/2}(F=2,m_F=0)$ transition,
obtained with 20 probe cycles for each value of the detuning. The linewidth of 10 Hz is approximately at the Fourier limit for the employed probe pulses of 90 ms duration.}
\end{figure}

Figure 1b shows a high-resolution excitation spectrum obtained with 90 ms long laser pulses, leading to an approximately Fourier-limited linewidth of
10 Hz, or a resolution $\Delta \nu/\nu$ of $1.4\cdot 10^{-14}$. Since the duration of the probe pulse is longer than the lifetime of the excited state (51 ms), the observed maximum excitation probability is limited by spontaneous decay.
In order to operate the system as a frequency standard, both wings of the resonance are probed alternately, and the probe light frequency is stabilized to the line center according to the difference of the measured excitation probabilities \cite{peik2}.

For quantitative studies of systematic frequency shifts we have compared the line center frequencies of two $^{171}$Yb$^+$ ions stored in separate traps \cite{schneider}.
The mean frequency difference between the two traps was in the sub-hertz range for nominally unperturbed conditions.
The dominant source of systematic uncertainty in the $^{171}$Yb$^+$ optical frequency standard is the so-called 
quadrupole shift of the atomic transition frequency. It is due to the interaction
of the electric quadrupole moment of the $D_{3/2}$ state with the gradient of static electric patch fields   
in the trap.
The frequency difference between the two systems shows 
a relative instability (Allan deviation) of $\sigma_y(1000\,{\rm s})=5\cdot 10^{-16}$, limited essentially by quantum projection noise \cite{peik2}. 

\begin{figure}[tbh]
\begin{center}
\includegraphics[width=8cm]{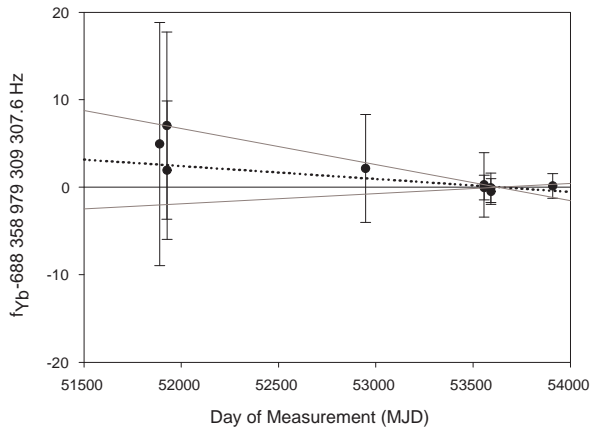}
\end{center}
\caption{Results of absolute frequency measurements of the Yb$^+$ optical frequency standard at 688 THz, plotted as a function of measurement date (MJD: Modified Julian Date), covering the period from December 2000 to June 2006. The dotted straight line is the result of a weighted linear regression through the data points. The gray lines indicate the $\pm 1 \sigma$ range for the slope as determined from the regression. The intercepts of the gray lines are chosen such that they cross in the point determined by the weighted averages of frequency and measurement date.}
\end{figure}

The absolute frequency of the $^{171}$Yb$^+$ standard at 688 THz was  measured relative to PTB's cesium fountain clock CSF1 \cite{csf1,csf2}.
The link between optical and microwave frequencies is established 
with a femtosecond-laser frequency comb generator \cite{udem}. 
The $^{171}$Yb$^+$ frequency was measured 
first in December 2000 \cite{Stenger1}. 
Fig. 2 shows the results of all frequency measurements performed up to June 2006, showing a steady decrease of the measurement uncertainty and excellent overall consistency.
Technical improvements over the years were made in the resolution of the ionic resonance, in the control of trapping conditions, and 
in the frequency comb generator setup. The latter was based on a titanium sapphire laser in the earlier experiments (until MJD 53000) and on an erbium doped fiber laser since \cite{adler}. Between July 2005 and June 2006 five absolute frequency measurements with continuous averaging times of up to 36~h were performed. The obtained statistical uncertainties (about $6\cdot10^{-16}$ after two days) are dominated by the white frequency noise of the cesium fountain. 
The relative systematic uncertainty contribution of the cesium reference of $1.2\cdot10^{-15}$ is based on a recently completed evaluation of CSF1 that is applicable to the 2005/06 measurements in retrospect. The systematic uncertainty contribution for the Yb$^+$ standard of $1.5\cdot10^{-15}$ is largely dominated by an estimate of 1~Hz for the stray-field induced quadrupole shift.

The weighted average over these measurements gives the present result for the frequency of the $^{171}$Yb$^+$ 
$^2S_{1/2}(F=0)\rightarrow {^2D}_{3/2}(F=2)$ transition: 
\begin{equation}
f_{\rm Yb}=688\, 358\, 979\, 309\, 307.6(1.4)~{\rm  Hz}
\end{equation}
with a total relative uncertainty of $2.0\cdot10^{-15}$.
This frequency value refers to a measurement at room temperature (296 K). Extrapolating to zero temperature (as it was done for the cesium clock) the frequency would have to be corrected for the AC Stark shift from blackbody radiation, i.e. increased by $0.37(5)$~Hz.

From a weighted linear regression of the sequence of frequency measurements (cf. Fig. 2) we obtain a slope of $(-0.54\pm0.97)$~Hz/yr, corresponding to 
a value for the fractional temporal variation of the frequency ratio $d\ln (f_{\rm Yb}/f_{\rm Cs})/dt=(-0.78\pm 1.40)\cdot 10^{-15}$~yr$^{-1}$, consistent with zero. This result can now be used to obtain new stringent limits on variations of $\alpha$ and $\mu$.

\bigskip
\noindent
{\bf 3. Limit on variations of the fine structure constant}
\medskip

For the analysis we use a simple parametrization that includes only a minimum of assumptions. 
The electronic transition frequency is expressed as \cite{karsh,pk}
\begin{equation}
f={\rm const}\cdot Ry \cdot F(\alpha)
\end{equation}
where 
$Ry=m_e e^4/(8\epsilon_0 h^3)\simeq 3.2898\cdot 10^{15}$ Hz 
is the Rydberg frequency, appearing as the common atomic scaling factor.
$F(\alpha)$ is a dimensionless function of $\alpha$
that takes relativistic contributions to the level energies into account \cite{prestage,dzu1}.
The numerical constant in front depends only on integer or half-integer quantum numbers characterizing the atomic structure
and is independent of time.  
The relative temporal derivative of the frequency $f$ can be written as:
\begin{equation}
\frac{d \ln f}{d t} =  \frac{d \ln Ry}{d t} + A \cdot \frac{d \ln \alpha}{d t}~~~~~{\rm with~}A\equiv \frac{d \ln F}{d \ln \alpha}.
\end{equation}       
The first term $d \ln Ry / d t$ represents a variation that would be common to all measurements of electronic transition frequencies: a change of the numerical value of the Rydberg frequency.\footnote{This statement refers to a dimensional quantity (frequency) in an SI unit (hertz). Since the SI second is fixed to a property of the $^{133}$Cs nucleus, possible variations of the quantity $Ry$ are not purely electromagnetic in origin -- as one would expect for an abstract physical quantity ``Rydberg constant'' -- but are also related to the strong interaction.}
The second term in Eq. 3 is specific to the atomic transition under study.
The  sensitivity factor $A$ for small changes of $\alpha$ has been calculated for the relevant transitions
by Dzuba, Flambaum {\it et al.} \cite{dzu1,dzu2}.         

To obtain a limit on $d \ln \alpha / d t$ we combine the data from $^{171}$Yb$^+$ with results 
on the constancy of a transition frequency in $^{199}$Hg$^+$, 
measured at NIST \cite{bize2,oskay}. 
The transition $5d^{10}6s~^2S_{1/2}\rightarrow 5d^96s^2~^2D_{5/2}$ of the mercury ion at 282~nm (1065 THz) with a natural linewidth of 1.9~Hz is studied. A sequence of measurements over the period from
August 2000 to November 2002
has resulted in a constraint on the fractional variation of  
$f_{\rm Hg}/f_{\rm Cs}$ at the level of $7\cdot 10^{-15}$~yr$^{-1}$ \cite{bize2}.
From more recent data\cite{oskay} with lower uncertainty a new limit  
$d \ln (f_{\rm Hg}/f_{\rm Cs})/dt=(0.27\pm 0.78)\cdot 10^{-15}$~yr$^{-1}$ can be derived.

The sensitivities of the ytterbium and mercury transition frequencies to changes of  $\alpha$ are quite different \cite{dzu1,dzu2}: $A_{\rm Yb}=0.88$ and $A_{\rm Hg}=-3.19$. 
The change of sign of $A$ between the two transitions reflects the fact that in Yb$^+$ a 6s-electron is excited to the empty 5d-shell, while in Hg$^+$ a hole is created in the filled 5d-shell when the electron is excited to 6s. The two measured drift rates together with Eq. 3 can now be used to calculate:
\begin{equation}
\frac{d\ln\alpha}{dt}= (-0.26\pm0.39)\cdot 10^{-15}~{\rm yr}^{-1},
\end{equation}
\begin{equation}
\frac{d\ln Ry}{dt}= (-0.55\pm1.11)\cdot 10^{-15}~{\rm yr}^{-1}.
\end{equation}
We conclude that to within an uncertainty of 4 parts in $10^{16}$ per year, the fine structure constant {\em is constant} in the present epoch.
This result can be qualified as model-independent because it is only the simple expression Eq. 2 and the {\it ab initio} atomic structure calculations of the $A$-values that have been used in the interpretation of the experimental data.
The constancy of $Ry$ implies that there is no temporal drift between atomic gross structure frequencies and the cesium hyperfine splitting. This is of importance for metrology because it means that a cesium clock and an optical atomic clock can be regarded as fundamentally equivalent.
  
\begin{figure}[tbh]
\begin{center}
\includegraphics[width=8cm]{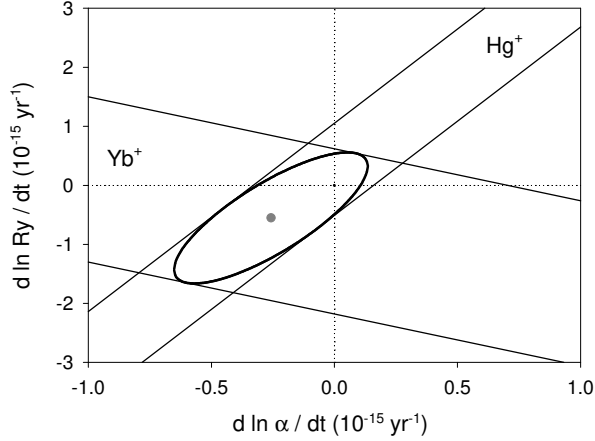}
\end{center}
\caption{Limits on temporal changes of the fine structure constant $\alpha$ and the Rydberg frequency $Ry$ obtained from the measured drift rates of frequencies in Yb$^+$ and Hg$^+$. Stripes mark the $1\sigma$ uncertainty of the individual measurements and the resulting combined standard uncertainty ellipse is shown within the central parallelogram.}
\end{figure}

Figure 3 shows how the two frequency drift rate measurements contribute to the constraints on changes in $\alpha$ and in $Ry$. Shown is also the standard uncertainty ellipse within the central parallelogram. In comparison to the earlier analysis that we published in 2004 \cite{peik}, based on the data available at the time, the $1\sigma$ uncertainty in $d\ln\alpha/dt$ has decreased by a factor of five.  Though over a very different time scale, the laboratory experiments have now reached the same  sensitivity for $d\ln\alpha/dt$ as the analysis of quasar absorption spectra, if a linear time evolution of $\alpha$ is assumed.

\bigskip
\noindent
{\bf 4. Limit on variations of the proton-to-electron mass ratio}
\medskip

The ratio $\mu=m_p/m_e$ is the second important dimensionless number whose possible variability was studied in astrophysical observations, especially in H$_2$ and OH molecular spectra. In the laboratory, molecular spectroscopy could also provide a direct access to 
a variability of $\mu$ \cite{schiller}, but has not yet reached the precision that has been achieved for electronic and hyperfine transitions in atoms. Data from the latter can, however, be used to derive a model-dependent limit on $d\ln \mu/dt$ along the lines described in 
Refs. [43,44].

In analogy to equation 3 the cesium atomic hyperfine transition frequency may be parametrized as
\begin{equation}
f_{\rm Cs}={\rm const}\cdot Ry \cdot \alpha ^2 \cdot \frac{\mu_{\rm Cs}}{\mu_B} \cdot F'(\alpha)
\end{equation}
where $F'(\alpha)$ describes relativistic corrections and $\mu_{\rm Cs}/\mu_B$ is the cesium nuclear magnetic moment in units of the Bohr magneton $\mu_B$.
Consequently, the logarithmic change of the ratio of an optical electronic frequency $f_{\rm opt}$ and $f_{\rm Cs}$ is given by
\begin{equation}
\frac{d \ln (f_{\rm opt}/f_{\rm Cs})}{d t} = -  \frac{d \ln (\mu_{\rm Cs}/\mu_B)}{d t} + 
(A_{\rm opt}-A'_{\rm Cs}-2)\,  \frac{d \ln \alpha}{d t},
\end{equation}       
where $A'_{\rm Cs}=d \ln F'_{\rm Cs}/d \ln \alpha=0.83$ \cite{dzu1}.
Linear regression of Eq. 7 with the data from the optical frequencies in Hg$^+$ and Yb$^+$  gives the result
\begin{equation}
\frac{d \ln (\mu_{\rm Cs}/\mu_B)}{d t} = (1.3 \pm 2.1)\cdot 10^{-15}~{\rm yr}^{-1}.
\end{equation}

A very precise result has been published on the constancy of the ratio of ground state hyperfine frequencies in $^{87}$Rb  with respect to $^{133}$Cs
\cite{bize}: $d\ln(f_{\rm Cs}/f_{\rm Rb})/dt=(0.05\pm0.53)\cdot 10^{-15}~{\rm yr}^{-1}$.
Since this frequency ratio scales like $(\mu_{\rm Cs}/\mu_{\rm Rb})\alpha^{0.49}$ one may
combine this experimental result with the limit on the variation of $\alpha$ (Eq. 4) in order  
to establish a limit on the temporal drift of the ratio of magnetic moments:
\begin{equation}
\frac{d \ln (\mu_{\rm Cs}/\mu_{\rm Rb})}{d t} = ( 0.18 \pm 0.56)\cdot 10^{-15}~{\rm yr}^{-1}.
\end{equation}

We can now derive model-dependent results on the proton $g$-factor $g_p$ and on the proton-to-electron mass ratio by using the Schmidt 
model \cite{blatt} that relates the nuclear magnetic moment to the magnetic moment of a single unpaired nucleon -- a proton in the cases of $^{87}$Rb and $^{133}$Cs.   
The predictions of the Schmidt model agree with the actual values to within about 25\% for 
$^{87}$Rb and to within 50\% for $^{133}$Cs. 
An uncertainty due to the application of this model is not included in the uncertainties of the limits presented in the following.
From the Schmidt model expression 
$\mu_{\rm Cs}/\mu_{\rm Rb}\propto (10-g_p)/(g_p+2)$,  
the limit on the drift of the magnetic moments can be applied to constrain a possible drift of $g_p$: 
\begin{equation}
\frac{d \ln g_p}{d t} = (- 0.09 \pm 0.28)\cdot 10^{-15}~{\rm yr}^{-1}.
\end{equation}

Using the Schmidt model expression for the cesium magnetic moment $\mu_{\rm Cs}=7(10-g_p)\mu_N/18$ (with the nuclear magneton $\mu_N$) and the fact that $\mu_B/\mu_N = m_p/m_e=\mu$ 
one can write the ratio of an optical
electronic transition frequency to the cesium hyperfine frequency as
\begin{equation}
\frac{f_{\rm opt}}{f_{\rm Cs}} \propto \frac{F_{\rm opt}}{\alpha^2 F'_{\rm Cs}}\cdot \frac{\mu}{10-g_p}.
\end{equation}
Combining the limit on the drift rate of $g_p$ (Eq. 10) with the results on the constancy of the optical frequencies in Hg$^+$ and Yb$^+$, a limit on the present temporal variation of the proton-to-electron mass ratio is obtained:
\begin{equation}
\frac{d \ln \mu}{d t} = (-1.2 \pm 2.2)\cdot 10^{-15}~{\rm yr}^{-1}.
\end{equation}

Compared to the previous evaluation \cite{drake} these limits to variations of the magnetic moments and mass ratio are now more strict by  about a factor of three. 
Like in the case of $\alpha$, the laboratory experiments reach the same sensitivity for $d\ln\mu/dt$ as the analysis of quasar absorption spectra under the assumption of a linear time evolution. We would like to point out, however, that the presented result on $d\alpha/dt$ should be valued higher than the one on $d\mu/dt$ because the latter is model-dependent. A more detailed parametrization of nuclear magnetic moments in terms of quark masses has been published \cite{vfted} and can be used if more precision data from different nuclei becomes available. 

\bigskip
\noindent
{\bf 5. Conclusion}
\medskip

The recent progress in the accuracy and reproducibility of optical frequency standards allows one to perform a sensitive laboratory search for temporal variations of fundamental constants. So far all these experimental results are consistent with the constancy of constants. We have presented an update 
on limits to variations based on recent optical frequency measurements in single trapped ions. Coincidentally, the laboratory experiments over time scales $\Delta t\approx5$~yr presently reach about the same sensitivity for changes of the fine structure constant $\Delta\alpha/\Delta t$ as the astrophysical observations in quasar absorption spectra over a time scale $\Delta t\approx 10^{10}$~yr. The laboratory search will soon become more sensitive, as the precision of frequency standards continues to improve. In addition, a larger variety of systems is now being investigated so that it will be possible to perform tests of the consistency if a first observation of a variation in the laboratory is reported.
The introduction of cooling techniques for small molecules will lead to more precise spectroscopy and to more sensitive tests for a variability of $\mu$. In nuclear physics, a very promising system has recently been identified in the closely spaced lowest states of the $^{229}$Th nucleus \cite{peik3}. This system has the potential to extend the scale for a search for changes in the strong and electromagnetic interactions by five orders of magnitude \cite{vf}.  

\bigskip
\noindent
{\bf Acknowledgments}
\medskip

We thank S. Karshenboim for many inspiring discussions.
This work was partly supported by Deutsche Forschungsgemeinschaft in SFB 407 and by grant RFP1-06-08 from the Foundational Questions Institute
(fqxi.org).

\end{document}